# Determining the Increase or Decrease of Photon Momentum: Dielectric or Plasmonic Mie Object at Air-Liquid Interface

M.R.C. Mahdy[1,2*], Hamim Mahmud Rivy[1], Ziaur Rahman Jony[1], Nabila Binte Alam[3], Nabila Masud[1], Ibraheem Muhammad Moosa[4] Chowdhury Mofizur Rahman[5], M Sohel Rahman[4*]

[1]*Department of Electrical & Computer Engineering, North South University, Bashundhara, Dhaka 1229, Bangladesh*

[2]*Pi Labs Bangladesh LTD, ARA Bhaban, 39, Kazi Nazrul Islam Avenue, Kawran Bazar, Dhaka 1215, Bangladesh*

[3] *Department of Computer Science & Engineering, Military Institute of Science and Technology, Dhaka, Bangladesh*

[4]*Department of Computer Science & Engineering, Bangladesh University of Engineering and Technology ECE Building, West Palasi, Dhaka-1205*

[5]*Department of Computer Science & Engineering, United International University, Dhaka, Bangladesh*

*Corresponding Authors: mahdy.chowdhury@northsouth.edu and msrahman@cse.Buet.ac.bd

**Though the interfacial tractor beam experiment supports the increase of photon momentum (i.e., Minkowski momentum), it is still a matter of investigation whether, inside matter, the photon momentum always increases. Considering the inhomogeneous or heterogeneous background, we have demonstrated that if the background and the half-immersed object are both non-absorbing, the transferred photon momentum to the object can be considered as the one of Minkowski *exactly at the interface*. In contrast, the presence of loss inside matter, either in the half-immersed object (i.e., plasmonic or lossy dielectric) or in the background, changes the whole situation. For such cases, our several demonstrations and proposed thought experiments have strongly supported the decrease of photon momentum instead of the usual perception of its increase. Although almost all the major radiation pressure experiments have so far supported the linear increase of photon momentum, our proposed simple experimental set-ups may introduce a novel way to observe and verify the exactly opposite proposal: the Abraham momentum of photon. Finally, as an interesting sidewalk, based on several parameters, a machine learning based system has been developed to predict the transferred momentum of photon within a very short time avoiding time-consuming full simulation.**

**Keywords:** Abraham-Minkowski controversy; dielectric interface; machine learning; optical force laws; optical pulling force; optical tractor beams.

## INTRODUCTION

In 1908, physicist Hermann Minkowski, proposed that the total momentum of an electromagnetic field inside any matter is equal to $\int \mathbf{D}\times\mathbf{B}\, dv$ [1]. This suggests that the photon momentum should actually increase and takes the value $p = n\hbar k_0$ according to quantum description. Here n is the refractive index of the medium, $k_0$ is the wave number of the electromagnetic wave in air and ħ is the reduced Planck constant. One year later, the German physicist Max Abraham proposed a different argument that the photon inside a medium would have a lower/ decreased momentum that is equal to $\int [(\mathbf{E}\times\mathbf{H})/c^2]\, dv$, or according to quantum description the value of photon momentum, $p = \hbar k_0/n$ [2]. Nearly 100 years later, there is still no clear answer as to which of these formulae is correct [3-5]. According to [5]: "both the Abraham and Minkowski forms of the momentum density are correct, with the former being the kinetic momentum and the later the canonical momentum." Notably, there is no simple experiment that supports the decrease of photon momentum (i.e., Abraham momentum of photon) inside a matter except for a few recent reports [6-8]. As a result, it would be certainly a great achievement if some very simple but realistic experiments are designed to measure the decrease of photon momentum inside a matter.

One of the recently popularized ideas in the area of photon momentum transfer or optical force is known as Tractor beam [9-11]. A tractor beam is a customized light beam that exerts a counter intuitive negative force on a scatterer [9-18], pulling it opposite to the propagation direction of light, in contrast to the conventional pushing force. Tractor beam experiments, which involve the material background [12-14], can also be investigated in details to understand the persistently debated roles of different photon momenta such as those of the Abraham-Minkowski controversy [3-5]. In this article, we make an attempt to investigate the Abraham-Minkowski controversy based on the tractor beam like effects.

We have investigated the light momentum transfer and related optical force on a scatterer (both absorbing and non-absorbing) floating on an interface of two or three different media (both absorbing and non-absorbing) as shown in Fig. 1(a), (b) and Fig. 2(a)-(d). Prior to this current work, for the interfacial tractor beam experiment, detailed calculations by ray tracing method and stress tensor equations showed that the pulling force is natural in air-water scheme due to the linear increase of photon momentum in the infinitely long water medium [12,14]. Interestingly, non-Minkowski formulations [14] show a pushing force contradicting the experimental observation in [12]. However, the suggested interpretation in favor of Minkowski photon momentum for interfacial tractor beam experiment has been questioned in [19]. Two different recent experiments have supported the observation of Abraham photon momentum for air-water interface [6, 7]. In addition, some theoretical works have suggested in favor of the possible existence of Abraham photon momentum [4, 5, 20-23] for different other situations.

Considering the inhomogeneous or heterogeneous background, we have theoretically demonstrated that if the background and the half-immersed object are both non-absorbing, the transferred photon momentum to the object can be considered as the one of Minkowski *exactly at the interface in time averaged scenario*. Remarkably, we have shown that even if the background's width is only a few nanometer (extremely small in comparison with the object), the half-immersed object experiences an optical pulling force. Several intuitive thought experiments have also been put forwarded to establish this fact.

In contrast, the presence of loss inside the matter, either in the half-immersed object (i.e., plasmonic or lossy dielectric) or in the background or in the extremely short background, changes the whole scenario. For such cases, our several demonstrations have strongly supported the decrease of linear momentum of photon instead of the usual perception of increase thereof. Though we have demonstrated several full wave simulation based results (along with analytical and qualitative explanations) in favor of linear decrease of photon momentum for distinct situations (i.e. absorption in the object or in the background or in the short background), we have concluded that: the easiest possible experiment to verify the decrease of photon momentum can be done by replacing the silicon object with a gold or a silver object in the well-known interfacial tractor beam experiment [12] and shining the usual Gaussian beam wave on it [12].

Finally, based on several parameters, a machine learning algorithm has been applied to predict the transferred momentum of photon at a shortest possible simulation time (instead of time consuming full wave simulations). Apart from our motivation of getting a quick prediction of the force type without a time-consuming full wave simulation, this artificial intelligence based investigation may open up a novel research avenue by providing us with yet another useful tool/approach to investigate other problems related to optical force and photon momentum inside a matter.

## METHODS

Throughout this paper, we refer to 'exterior' ('interior') or 'outside' ('inside') forces as those evaluated outside (inside) the volume of the macroscopic objects. We have done all the full wave electromagnetic simulations using COMSOL MULTI PHYSICS software [24].

The proposed set-ups are illustrated in Figs. 1 (a)-(c). The source is a simple $z$ -polarized (cf. Fig. 1) plane wave $\mathrm{H}=\mathrm{H}_0 e^{j\beta(x\cos\theta - y\sin\theta)}$ (the varying incident angle $\theta$ is with the '$-x$' axis), where $\mathrm{H}_0 = 1v/m$ and the wavelength is always 632.5 nm. The 'outside optical force' [25, 26] is calculated by the integration of time averaged external Minkowski [27-31] stress tensor at r=$a^+$ employing the background fields of the scatterer of radius *a* as follows [in this article, all time averaged forces have been calculated based on this equation]:

$$\begin{aligned} \left\langle \boldsymbol{F}_{\text{Total}}^{\text{Out}} \right\rangle &= \oint \left\langle \overline{\overline{\boldsymbol{T}}}^{\text{out}} \right\rangle \cdot d\boldsymbol{s} \\ \left\langle \overline{\overline{\boldsymbol{T}}}^{\text{out}} \right\rangle &= \frac{1}{2} \text{Re}[\boldsymbol{D}_{\text{out}} \boldsymbol{E}_{\text{out}}^* + \boldsymbol{B}_{\text{out}} \boldsymbol{H}_{\text{out}}^* - \frac{1}{2} \overline{\overline{\boldsymbol{I}}} (\boldsymbol{E}_{\text{out}}^* \cdot \boldsymbol{D}_{\text{out}} + \boldsymbol{H}_{\text{out}}^* \cdot \boldsymbol{B}_{\text{out}})] \end{aligned} \quad (1)$$

Here 'out' represents the exterior total field of the scatterer; $\boldsymbol{E}$ , $\boldsymbol{D}$ , $\boldsymbol{H}$ and $\boldsymbol{B}$ are the electric field, electric displacement field, magnetic field and magnetic induction field respectively, $\langle \ \rangle$ represents the time average and $\overline{\overline{\boldsymbol{I}}}$ is the unity tensor. In contrast, the 'internal optical force' of Minkowski [27, 32] is calculated by the integration of time averaged Minkowski stress tensor at r=$a^-$ employing the interior fields of the scatterer of radius *a* as follows:

$$\begin{aligned} \left\langle \boldsymbol{F}_c^{\text{in}} \right\rangle = \left\langle \boldsymbol{F}_{\text{Mink.}}^{\text{Bulk}} \right\rangle (\text{in}) &= \oint \left\langle \overline{\overline{\boldsymbol{T}}}^{\text{in}} \right\rangle \cdot d\boldsymbol{s} \\ \left\langle \overline{\overline{\boldsymbol{T}}}^{\text{in}} \right\rangle &= \frac{1}{2} \text{Re}[\boldsymbol{D}_{\text{in}} \boldsymbol{E}_{\text{in}}^* + \boldsymbol{B}_{\text{in}} \boldsymbol{H}_{\text{in}}^* - \frac{1}{2} \overline{\overline{\boldsymbol{I}}} (\boldsymbol{E}_{\text{in}}^* \cdot \boldsymbol{D}_{\text{in}} + \boldsymbol{H}_{\text{in}}^* \cdot \boldsymbol{B}_{\text{in}})] \end{aligned} \quad (2)$$

Here 'in' represents the interior field of the scatterer. Notably, the 'internal optical force' of Minkowski represents the total conducting force (due to the interaction of the photon with the free charges inside the absorbing objects) inside an object. For lossless objects, Eq (2) always yields zero force inside the objects [27, 32].

## RESULTS AND DISCUSSIONS

Though the interfacial tractor beam experiment supports the linear increase of photon momentum (i.e. Minkowski momentum), the following question is still a matter of investigation: inside the matter, does the photon momentum always increases; and even if it increases, does it increase throughout the whole medium or just at the interface? In [33], it has been argued that the increase of linear momentum usually occurs at the interface of two different media due to the reduced impedance mismatch. But no conclusive proof is still presented regarding this very interesting argument given in [33]. Based on a few proposed 'thought experiments', we shall first check this proposal for interfacial tractor beam set-up at first.

From Fig. 1 (a), it is evident that if we increase the refractive index of the second medium, the magnitude of optical pulling force increases which supports the Minkowski photon momentum. But Fig 1(b) reveals that if we consider extremely small lower background (water) medium [around 10 nm width, where object diameter is 4000 nm], still an optical pulling force occurs. So, the main point is that even for an extremely short background, the proposition of the linear momentum increase remains valid. Now, the obvious question arises whether there is any alternative way to verify the robustness of such a proposal that claims an increase of photon momentum just at the interface of two different media. This is addressed next along with few other distinct issues.

In Fig. 2 (a) and (b), we have checked the veracity of the proposition of such linear momentum increases at the interface of two media. In Fig. 2(a), it has been shown that if absorption is introduced in the infinite background, optical pulling force vanishes for the half-immersed object. The effect of absorption in the background medium is discussed later. Here we have just used the fact that absorption in the background can destroy the optical pulling force and it may also destroy the increase of linear photon momentum. Now, if we again introduce the extremely short non-absorbing water background (width of 30nm where the object diameter is 4000 nm) in between the scatter and that absorbing infinite background, again the optical pulling force is observed in Fig. 2(b). This provides clear evidence in favor of the increase of photon momentum exactly at the interface of two different media when it is non-absorbing.

So, we can conclude that: the touching short background plays a vital role on time averaged optical force/transferred momentum of photon under certain specific conditions (i.e., the object should be a Mie object). This fact can also be intuitively guessed by a simple way: by putting a Silicon object such that there would be an extremely small air gap of only 2nm (i.e., short background width and it is just air) between the lower long background and the object in APPENDIX A. Even if the refractive index of the lower long background is increased up to the value of 1.87 (one of the maximum values of ref. index possible for liquid materials [34]), the force is always found to be a pushing force based on full wave simulations as shown in APPENDIX A [cf. Fig. 1A in APPENDIX A]. In contrast, even if the lower long background is air but the short background (width around 10 nm) is just water, the force is found to be a pulling force in Fig. 1(b) in this article [also cf. Fig. 2(b) for another case]. So, the short background is indeed an imperative criterion which plays a significant role in such set-ups.

Although for the non-absorbing object and non-absorbing background, the transferred momentum of photon usually increases, this scenario changes when applied in the presence of absorption; either in the sub-merged object or in the embedding lower background. At first, we can check the case of half immersed absorbing object. If we replace a half immersed non-absorbing dielectric by a lossy dielectric or a plasmonic object, it is clearly observed that the pulling force fully vanishes as shown in Fig. 2(c); this suggests the linear decrease of the transferred photon momentum. Finally, if we consider a short absorbing background (instead of an infinitely long one) with a half immersed lossless dielectric object, then again optical pulling force vanishes as shown in Fig. 2(d) supporting the decrease of photon momentum. Next, we address the obvious question whether there exists any alternative way to investigate the veracity of the proposal of the decrease of photon momentum in an absorbing medium.

A simple fact is that, if the upper medium is a high refractive indexed lossless medium but the lower one is air, then the lossless object experiences an optical pushing force as shown in Fig. 3(a) [as the transferred momentum from upper medium to the air medium decreases a lot which assists optical pushing force]. But to identify whether the presence of absorption really affects the decrease of photon momentum or not, we shall introduce some absorption in the upper medium and conduct a thought experiment. Suppose a lossless silicon object is half immersed where the upper medium has high refractive index along with moderate loss and the lower medium is air [cf. Fig. 3(b)]. If the photon momentum in the above medium is lower than the air medium, then due to the increase of refractive index of that medium (which should decrease the photon momentum according to Abraham's definition: $p = \hbar k_0/n$), there should be a very good chance of observing a pulling force instead of pushing force. According to our full wave simulation result, it really happens as shown in Fig. 3(b). But it should be noted that, if the refractive index of the upper medium is not very high (with moderate imaginary part) such pulling force may not occur.

However, the most challenging part is the availability of a liquid having a very high refractive index along with a moderate loss. In contrast, the simpler experiments, proposed in Figs. 2(a), (c) and (d), can be conducted very easily, which also support that the linear momentum of photon decreases in the presence of absorption. The easiest possible experiment to verify the decrease of photon momentum can be done by replacing the silicon object with a gold or a silver object in the well-known interfacial tractor beam experiment [12] and shining the usual Gaussian beam wave on it [12]. The findings and demonstrations in this article suggest that optical pushing force should be observed instead of optical pulling force in such a set-up.

## Explanation of the observed results

Now the first question is what actually happens during the decrease of photon momentum [Figs 2 and 3]. And the second question is what the possible differences between the non-absorbing situations and the absorbing cases are. Now, we shall investigate the idea more intuitively.

It is well known that Minkowski stress tensor [27-31, 35-37] has divergence free nature, which suggests that, if the Minkowski stress tensor is applied inside a non-absorbing object employing the internal field of the embedded object, it would lead to zero time-averaged total force. The difference between the internal Minkowski stress tensor and the external Minkowski stress tensor, exactly at the object and background interface, leads to the Helmholtz's surface force. The time averaged pulling force observed in the interfacial tractor beam experiment can be explained/ calculated solely based on this Helmholtz's surface force given in forthcoming Eq (5) [which appears just at the interface of two different media (cf. Fig 10 (a), (b) and Fig 12 (a) given in ref. [38])] and the linear increase of the photon momentum occurs *exactly at the interface* of two different media [Helmholtz/ Minkowski force also suggests so]. But when absorption is introduced inside the half-immersed object, the force inside the absorbing object has a non-zero value equal to Eq (2) [cf. ref [27] for more detail on Eq (2)]. According to [27,32], that same time averaged conduction force $\left\langle \boldsymbol{F}_c^{\text{in}} \right\rangle$ in Eq (2) (that arises due to the free currents inside the scatterer) can also be written as volumetric force method [17, 27, 32, 39]:

$$\left\langle \boldsymbol{F}_c \right\rangle = \int \left\langle \boldsymbol{f}_c \right\rangle dv = \int \frac{1}{2} \text{Re}\left[ \omega \varepsilon_I \boldsymbol{E}_{\text{in}} \times \boldsymbol{B}_{\text{in}}^* - \omega \mu_I \boldsymbol{H}_{\text{in}} \times \boldsymbol{D}_{\text{in}}^* \right] dv \quad (3)$$

Here $\varepsilon_I$ and $\mu_I$ are the imaginary parts of the permittivity and permeability of a scatterer respectively. It should be noted that, if there is no absorption inside an object (i.e. lossless object), the total internal force $\left\langle \boldsymbol{F}_c^{\text{in}} \right\rangle$ becomes zero which can be understood from Eq (3) [as there would be no $\varepsilon_I$ and $\mu_I$ in Eq (3)].

The connection between Eq (1) and Eq (2) can be written as:

$$\int \left\langle \overline{\overline{\boldsymbol{T}}}_{\text{Mink.}} (\text{out}) \right\rangle \cdot d\boldsymbol{s} = \left\langle \boldsymbol{F}^{\text{Total}} \right\rangle = \left\langle \boldsymbol{F}_{\text{Mink.}}^{\text{Bulk}} \right\rangle (\text{in}) + \left\langle \boldsymbol{F}_{\text{Mink.}}^{\text{Surface}} \right\rangle \quad (4)$$

$\left\langle \boldsymbol{F}_{\text{Mink.}}^{\text{Surface}} \right\rangle$ is known as the time averaged surface force of Minkowski force law which is well-known as Helmholtz's surface force [40] (its independent microscopic derivation is available in [41]):

$$\boldsymbol{f}_{\text{Helmholtz}}^{\text{Surface}} = \left[ -\frac{1}{2} \boldsymbol{E}^2 \Delta \varepsilon - \frac{1}{2} \boldsymbol{H}^2 \Delta \mu \right]_{r=a} \quad (5)$$

According to Eq (4), on the boundary *r=a* of any object, the quantity: $[\overline{\overline{\boldsymbol{T}}}(\text{out}) - \overline{\overline{\boldsymbol{T}}}(\text{in})] \cdot \hat{\boldsymbol{n}} = \boldsymbol{f}_1^{\text{Surface}}$ [ $\hat{\boldsymbol{n}}$ being the local unit outward normal of the object surface] should be exactly the surface force density of Eq (5). In APPENDIX B, after doing some detail analytical calculations we have shown that we get exactly Eq (5), the surface force law of Helmholtz, at the object boundary from the quantity $[\overline{\overline{\boldsymbol{T}}}(\text{out}) - \overline{\overline{\boldsymbol{T}}}(\text{in})] \cdot \hat{\boldsymbol{n}} = \boldsymbol{f}_1^{\text{Surface}}$ .

-Due to this loss of momentum [Eq (2) or Eq (3)] inside the absorbing object, the ultimate momentum of photon at the interface of the two different media (i.e. the half immersed object and the second background) decreases. For the other case, when the object is non-absorbing but the second medium (or the extremely short background) is absorbing, this same thing happens (non-zero value of ‘internal’ Minkowski stress tensor inside the absorbing medium [given in Eq (2)]) and the momentum of photon decreases.

In [42], the connection of optical pulling force due to the incidence of Bessel beam has been discussed based on the singular points of Poynting vector distribution. For interfacial tractor beam set-up with non-absorbing cases, it is observed that whenever optical pulling force occurs, the magnitude of Poynting vector is concentrated just at the right interface of the sub-merged object and the connected/touching lower background [cf. Figs 1(a) and (b)]. Interestingly, the situation fully changes for the absorbing case: either the absorbing object or the absorbing background. In the presence of loss (and hence the optical pushing force), usually the high concentration of Poynting vector appears in the upper air medium above the interface region [cf. Figs 2 (a) and (c)]. But a notable exception occurs for the case of extremely short background with absorption, where optical pushing force occurs even though the higher concentration of Poynting vector appears at the lower interface (at the boundary of the object and that short background [cf. Fig. 2(d)]).

**Application of Machine Learning Techniques for predicting the type of optical force**

Finally, we take an interesting sidewalk from our primary focus and investigate whether machine learning based data analytics can provide us an accurate prediction on the transferred momentum of photon for the more generic cases based on the discussed complex set-ups in this article. The motivation behind this rather unusual investigation is as follows. A lot of parameters actually contribute in determining the optical force type (i.e., pulling vs. pushing) and a full wave simulation [24] takes long time. In contrast, a quick accurate result in this regard could be extremely useful for experimental physicists at least to make quick primary decisions in many situations. So, we have made an attempt to develop a machine learning based system, using WEKA workbench [43], which can accurately predict the outcome extremely faster. However, the accuracy of this prediction is naturally dependent on the dataset used for training the system. The dataset has been prepared based on some full wave simulations and is provided in Chart 1s in the supplement**.**

We have used the WEKA workbench [43] to develop the machine learning based classifiers and to conduct the computational experiments. The data given in the supplement (Chart 1s of the supplement) has been used to train a system that is able to predict the optical force type. In what follows, this data will be referred to as the training dataset. The training dataset in Chart 1s has been prepared based on several full wave simulations. The varying parameters used in the dataset are: particle radius/lambda, particle loss, refractive index of: the particle, background and short background; loss at: long background and short background; and short background width. Thus Chart 1s in the supplement article has the first 8 columns for the parameters with the last (i.e., 9$^{th}$) column representing the time averaged force experienced by the particle; here '1' ('0') represents optical pulling (pushing) force. It should be noted that *we have varied the incident angle of light from 0 degree to 90 degree; if the time averaged force is negative for most of the angles, we have defined it as '1'; and in contrast, if the time averaged force is positive for most of the angles, we have defined it as '0'.*

Optical pulling force ('1') suggests that: the linear momentum of photon increases just in the touching background in time averaged scenario. In contrast, optical pushing force ('0') suggests the decrease of photon momentum in time averaged scenario.

For our classification task, we have used some distinct classifiers, namely, Naïve Bayes [44-46], Simple Logistics [44, 46], J48 [44, 48], Random Forest [44, 49], Random Tree [44, 49] and so on. Following the machine learning literature, we have employed 10-fold cross validations: the training dataset is randomly partitioned into 10 equal sized subsets; of the 10 subsets, a single subset is retained as the validation data for testing the model, and the remaining 9 subsets are used as training data. The cross-validation process is then repeated 10 times, with each of the 10 subsets used exactly once as the validation data. The 10 results can then be averaged to produce a single estimation. Table 1 reports the accuracy measures found after performing 10-fold cross validations for each of the classifiers. From the results reported in Table 1, J48, Random Forest and Random Tree classifiers have predicted the final results with the highest accuracy (more than 95%, which is quite satisfactory).

Table 1: 10-fold cross validation results for different classifiers

| Sl. | Classifiers | Accuracy |
|---|---|---|
| 1 | **Naïve Bayes** | 74.12% |
| 2 | **Simple Logistic** | 83.07% |
| 3 | **Decision table** | 92.02% |
| 4 | **Jrip** | 95.13% |
| 5 | **J48** | 96.49% |
| 6 | **Random Forest** | 97.27% |
| 7 | **Random Tree** | 95.71% |

So, given the values of the required parameters, our machine learning based predictor tool can be used to accurately and very quickly predict the force type. Apart from our motivation of getting a quick prediction of the force type without a time-consuming full wave simulation, this artificial intelligence based investigation may open up a novel research avenue by providing us with yet another useful tool/approach to investigate other problems related to optical force and photon momentum inside a matter.

## CONCLUSION

In this work, considering the inhomogeneous or heterogeneous background, we have theoretically demonstrated that if the background and the half immersed object are both non-absorbing, the transferred photon momentum to the object can be considered as the one of Minkowski *exactly at the interface* (in time averaged scenario). Remarkably, we have shown that even if the background's width is only a few nanometer (extremely small in comparison with the object), the half-immersed object experiences an optical pulling force. In contrast, the presence of loss inside matter, either in the half immersed object (i.e. plasmonic or lossy dielectric) or in the background changes the whole situation. For such cases, our several demonstrations and proposed thought experiments have strongly supported the decrease of photon momentum instead of the usual perception of its increase. Finally, based on several parameters, a machine learning based technique has been applied to quickly predict the transferred momentum of photon, which may open up a novel research avenue by providing a useful tool/approach (i.e. artificial intelligence) to investigate other problems related to optical force and photon momentum inside a matter.

# Figures and Captions List

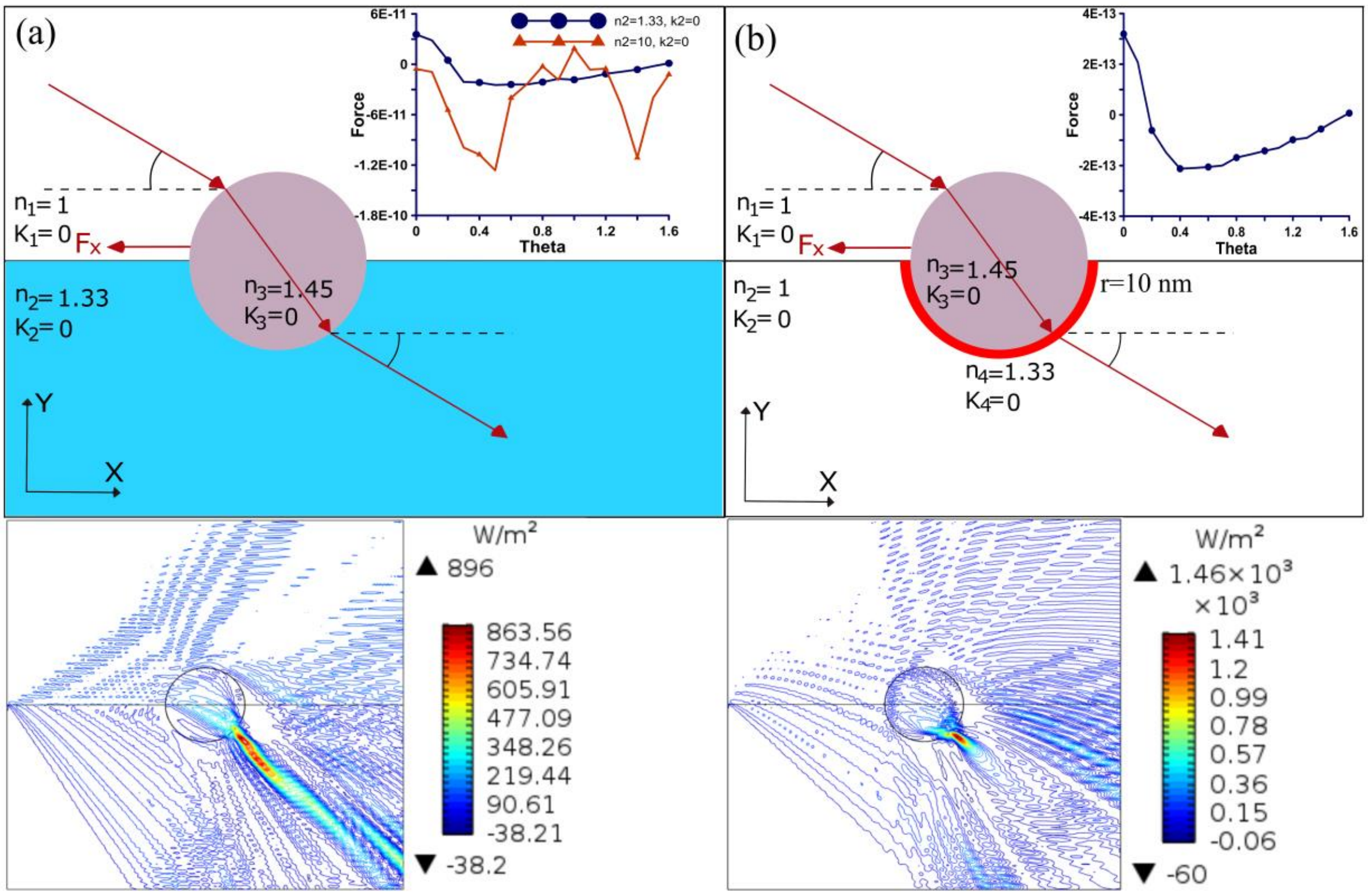

Fig. 1 Diameter of the Silicon object is 4000 nm and time averaged force has been calculated using Eq (1). The lower figure shows the Poynting vector distribution for the incident angle of 45 degree for each specific case. (a) Object half immersed in air and water medium of ref. index 1.33 (or a medium with refractive index 10). Inset: optical pulling force occurs for different angles of incident light. For the case of ref. index=10, Optical pulling force increases at least around100 times than air-water case for different angles of incident light. Lower Fig: Poynting vector distribution for the case- the second background is water. (b) Object half immersed in air and extremely small (width of 10 nm only) background of water. Inset: optical pulling force occurs (almost similar to the force of the air-water case in Fig. 1(a)) for different angles of incident light.

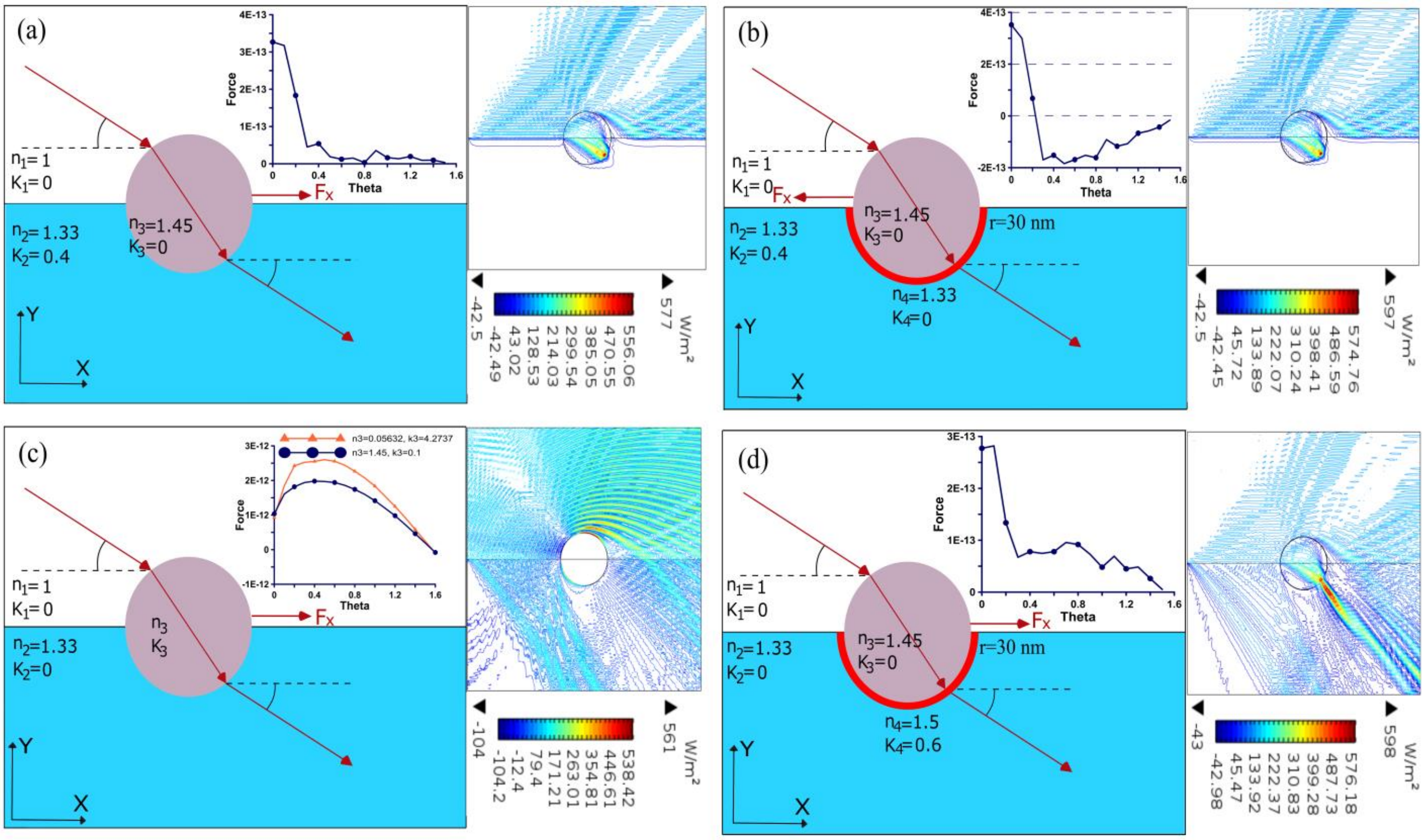

Fig. 2 Diameter of the Silicon object is 4000 nm and time averaged force has been calculated using Eq (1). Right side inset in all figures: the Poynting vector distribution for the incident angle of 45 degree. (a) Object half immersed in air and lossy water (absorption coefficient value is 0.4). Inset: optical pushing force occurs for different angles of incident light. (b) Object half immersed in air and lossy water (absorption coefficient value is 0.4). But a third background (small lossless water background with its width of 30 nm) is introduced between the object and the long absorbing water background. Inset: optical pulling force occurs for different angles of incident light. (c) (i) An absorbing Silicon object [absorbing coefficient 0.1; lossy dielectric] or (ii) a plasmonic Silver object -half immersed in air and non-absorbing water. Inset: optical pushing force for different angles of incident light for both the cases. Right side inset: Poynting vector (incident angle of 45 degree for plasmonic object [the absorbing silicon case is also very similar; hence not shown here]). (d) A lossless Silicon object half immersed in air and lossless water. Small lossy background's width: 30 nm (absorption coefficient value is 0.6 and refractive index 1.5). Inset: optical pushing force occurs for different angles of incident light.

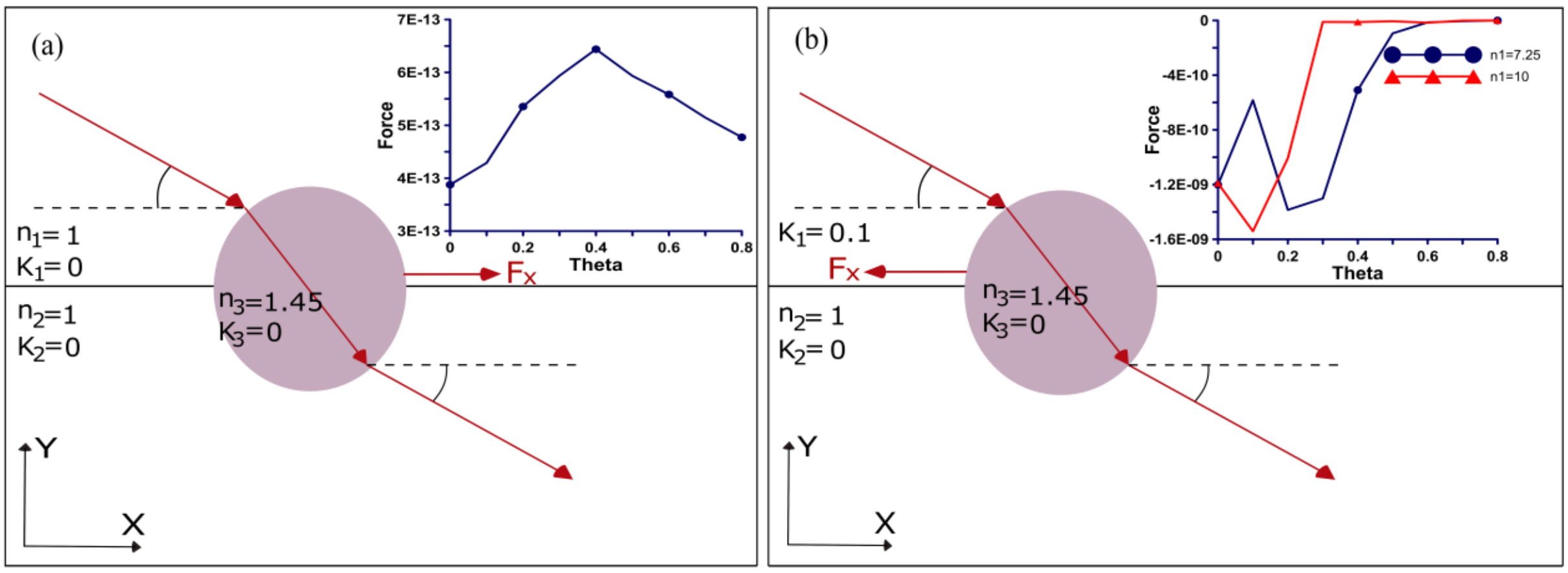

Fig. 3 Diameter of the Silicon object is 4000 nm and time averaged force has been calculated using Eq (1). (a) Object half immersed in a hypothetical lossless upper medium of refractive index 7.25 and the lower medium of air. Inset: optical pushing force occurs for different angles of incident light. (b) Object half immersed in two distinct hypothetical lossy upper media of: refractive index (i) 7.25 or (ii) 10 (absorption coefficient always 0.1) but the lower medium is always air. Inset: optical pulling force occurs for different angles of incident light for both the cases.

**APPENDIX A**:

**When short background is air instead of a material medium**

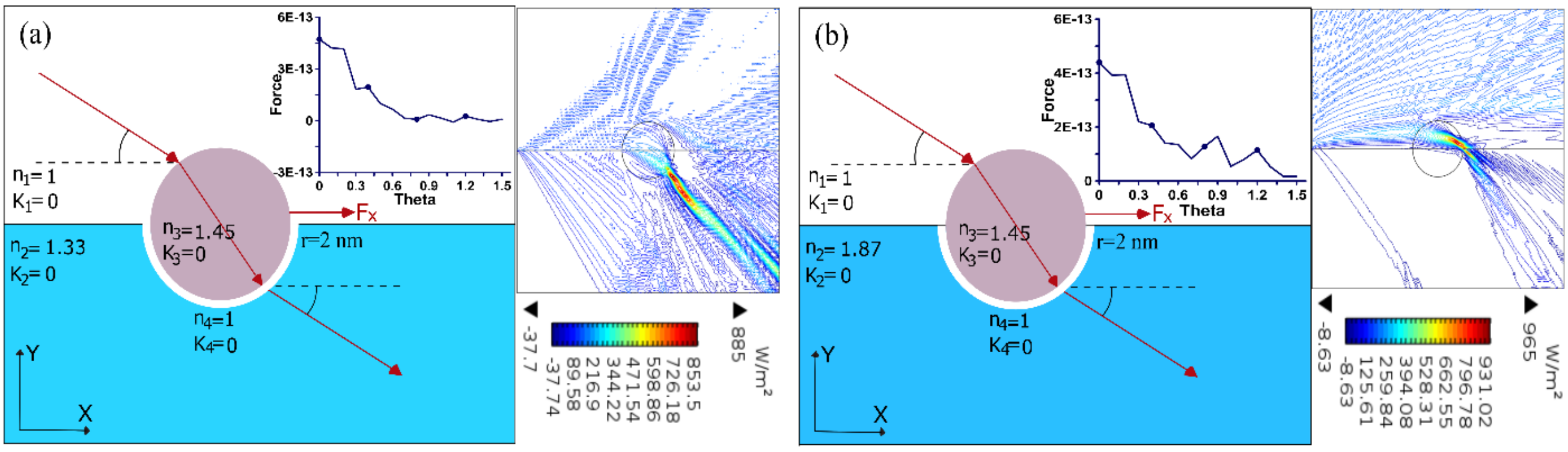

Fig. 1A Diameter of the Silicon object is 4000 nm and time averaged force has been calculated using Eq (1) given in main article. The width of the hypothetical short background of air is 2 nm. Time averaged optical force is found always pushing when: (a) Long background is water and (b) long background has refractive index 1.87.

**APPENDIX B:**

**Derivation of Helmholtz surface force from Minkowski stress tensor**

The 'outside optical force' is calculated by the integration of time averaged external Minkowski stress tensor at r=$a^+$ employing the background fields of the scatterer of radius $a$:

$$\begin{aligned}\left\langle \boldsymbol{F}_{\text{Total}}^{\text{Out}} \right\rangle &= \oint \left\langle \overline{\overline{\boldsymbol{T}}}^{\text{out}} \right\rangle \cdot d\boldsymbol{s} \\ \left\langle \overline{\overline{\boldsymbol{T}}}^{\text{out}} \right\rangle &= \frac{1}{2}\text{Re}[\boldsymbol{D}_{\text{out}}\boldsymbol{E}_{\text{out}}^{*} + \boldsymbol{B}_{\text{out}}\boldsymbol{H}_{\text{out}}^{*} - \frac{1}{2}\overline{\overline{\boldsymbol{I}}}(\boldsymbol{E}_{\text{out}}^{*}\cdot\boldsymbol{D}_{\text{out}} + \boldsymbol{H}_{\text{out}}^{*}\cdot\boldsymbol{B}_{\text{out}})]\end{aligned} \quad . \tag{1B}$$

Here 'out' represents the exterior total field of the scatterer; $\boldsymbol{E}$ , $\boldsymbol{D}$ , $\boldsymbol{H}$ and $\boldsymbol{B}$ are the electric field, electric displacement field, magnetic field and magnetic induction field respectively, $\langle\ \rangle$ represents the time average and $\overline{\overline{\boldsymbol{I}}}$ is the unity tensor.

On the other hand, the 'internal optical force' of Minkowski is calculated by the integration of time averaged Minkowski stress tensor at r=$a^-$ employing the interior fields of the scatterer of radius $a$:

$$\left\langle \boldsymbol{F}_c^{\text{in}} \right\rangle = \left\langle \boldsymbol{F}_{\text{Mink.}}^{\text{Bulk}} \right\rangle (\text{in}) = \oint \left\langle \overline{\overline{\boldsymbol{T}}}^{\text{in}} \right\rangle \cdot d\boldsymbol{s}$$

$$\left\langle \overline{\overline{\boldsymbol{T}}}^{\text{in}} \right\rangle = \frac{1}{2}\text{Re}[\boldsymbol{D}_{\text{in}}\boldsymbol{E}_{\text{in}}^* + \boldsymbol{B}_{\text{in}}\boldsymbol{H}_{\text{in}}^* - \frac{1}{2}\overline{\overline{\boldsymbol{I}}}(\boldsymbol{E}_{\text{in}}^* \cdot \boldsymbol{D}_{\text{in}} + \boldsymbol{H}_{\text{in}}^* \cdot \boldsymbol{B}_{\text{in}})] . \quad (2B)$$

Here 'in' represents the interior field of the scatterer; $\boldsymbol{E}$ , $\boldsymbol{D}$ , $\boldsymbol{H}$ and $\boldsymbol{B}$ are the electric field, electric displacement field, magnetic field and magnetic induction field respectively.

Now, from the non-diagonal (ND) components of Minkowski Stress tensors given in Eq (1B) and (2B), we get:

$$[\overline{\overline{\boldsymbol{T}}}_{\text{Mink}}^{\text{MIX}}(\text{out}) - \overline{\overline{\boldsymbol{T}}}_{\text{Mink}}^{\text{MIX}}(\text{in})] \cdot \hat{\boldsymbol{n}}\;_{r=a} = \begin{bmatrix} [\epsilon_b \boldsymbol{E}_{\text{out}}^{\perp} \cdot \boldsymbol{E}_{\text{out}}^{\perp} - \epsilon_s \boldsymbol{E}_{\text{in}}^{\perp} \cdot \boldsymbol{E}_{\text{in}}^{\perp}]\hat{\boldsymbol{n}} + [\epsilon_b \boldsymbol{E}_{\text{m out}}^{\parallel} \cdot \boldsymbol{E}_{\text{out}}^{\perp} - \epsilon_s \boldsymbol{E}_{\text{m in}}^{\parallel} \cdot \boldsymbol{E}_{\text{in}}^{\perp}]\hat{\boldsymbol{m}} \\ +[\epsilon_b \boldsymbol{E}_{\text{q out}}^{\parallel} \cdot \boldsymbol{E}_{\text{out}}^{\perp} - \epsilon_s \boldsymbol{E}_{\text{q in}}^{\parallel} \cdot \boldsymbol{E}_{\text{in}}^{\perp}]\hat{\boldsymbol{q}} \end{bmatrix}_{r=a}$$
$$+ \begin{bmatrix} [\mu_b \boldsymbol{H}_{\text{out}}^{\perp} \cdot \boldsymbol{H}_{\text{out}}^{\perp} - \mu_s \boldsymbol{H}_{\text{in}}^{\perp} \cdot \boldsymbol{H}_{\text{in}}^{\perp}]\hat{\boldsymbol{n}} + [\mu_b \boldsymbol{H}_{\text{m out}}^{\parallel} \cdot \boldsymbol{H}_{\text{out}}^{\perp} - \mu_s \boldsymbol{H}_{\text{m in}}^{\parallel} \cdot \boldsymbol{H}_{\text{in}}^{\perp}]\hat{\boldsymbol{m}} \\ +[\mu_b \boldsymbol{H}_{\text{q out}}^{\parallel} \cdot \boldsymbol{H}_{\text{out}}^{\perp} - \mu_s \boldsymbol{H}_{\text{q in}}^{\parallel} \cdot \boldsymbol{H}_{\text{in}}^{\perp}]\hat{\boldsymbol{q}} \end{bmatrix}_{r=a} \quad (3B)$$

Here, 'MIX' represents the mixed diagonal and non-diagonal elements of the stress tensor, which are not connected with the identity tensor, $\overline{\overline{\boldsymbol{I}}}$ . $\varepsilon_{\text{b}}$ and $\mu_{\text{b}}$ are fixed background permittivity and permeability, and $\varepsilon_s$ and $\mu_s$ are fixed permittivity and permeability of the scatterer. 'out' represents the total fields (incident plus scattered field) outside a scatterer. 'in' represents the fields inside a scatterer. Electric field at the object and background boundary is defined as: $\boldsymbol{E} = \boldsymbol{E}_n^{\perp}\hat{\boldsymbol{n}} + \boldsymbol{E}_{\text{m}}^{\parallel}\,\hat{\boldsymbol{m}} + \boldsymbol{E}_{\text{q}}^{\parallel}\,\hat{\boldsymbol{q}}$ where $\hat{\boldsymbol{n}}$ , $\hat{\boldsymbol{m}}$ and $\hat{\boldsymbol{q}}$ are mutually orthogonal arbitrary unit vectors, which are applicable for different co-ordinate systems such as Cartesian or Spherical or Cylindrical. $\hat{\boldsymbol{n}}$ is the local unit normal of the object surface, which is considered aligned towards the direction of wave vector direction (for simplicity). $\boldsymbol{E}^{\parallel}$ and $\boldsymbol{E}^{\perp}$ are the parallel and perpendicular components of electric fields at the background and object boundary. In a very similar way, the magnetic field has also been defined. Now, by employing the electromagnetic boundary conditions in above Eq (3B):

$$[\overline{\overline{\boldsymbol{T}}}_{\text{Mink}}^{\text{MIX}}(\text{out}) - \overline{\overline{\boldsymbol{T}}}_{\text{Mink}}^{\text{MIX}}(\text{in})] \cdot \hat{\boldsymbol{n}}\;_{r=a} = [\epsilon_b \boldsymbol{E}_{\text{out}}^{\perp} \cdot \boldsymbol{E}_{\text{out}}^{\perp} - \epsilon_s \boldsymbol{E}_{\text{in}}^{\perp} \cdot \boldsymbol{E}_{\text{in}}^{\perp}]\hat{\boldsymbol{n}} + [\mu_b \boldsymbol{H}_{\text{out}}^{\perp} \cdot \boldsymbol{H}_{\text{out}}^{\perp} - \mu_s \boldsymbol{H}_{\text{in}}^{\perp} \cdot \boldsymbol{H}_{\text{in}}^{\perp}]\hat{\boldsymbol{n}} \quad (4B)$$

In contrast, from the pure diagonal (D) components we get [after employing the electromagnetic boundary conditions]:

$$[\bar{\bar{\boldsymbol{T}}}_{\text{Mink}}^{\text{D}}(\text{out}) - \bar{\bar{\boldsymbol{T}}}_{\text{Mink}}^{\text{D}}(\text{in})] \cdot \hat{\boldsymbol{n}}_{r=a} =$$

$$-\frac{1}{2}\left[\begin{array}{l}\left(\epsilon_b \boldsymbol{E}_{\text{out}}^{\perp} \cdot \boldsymbol{E}_{\text{out}}^{\perp} - \epsilon_s \boldsymbol{E}_{\text{in}}^{\perp} \cdot \boldsymbol{E}_{\text{in}}^{\perp}\right)\hat{\boldsymbol{n}} + \left(\epsilon_b \boldsymbol{E}_{\text{m out}}^{\parallel} \cdot \boldsymbol{E}_{\text{m out}}^{\parallel} - \epsilon_s \boldsymbol{E}_{\text{m in}}^{\parallel} \cdot \boldsymbol{E}_{\text{m in}}^{\parallel}\right)]\hat{\boldsymbol{m}} \\ + \left(\epsilon_b \boldsymbol{E}_{\text{q out}}^{\parallel} \cdot \boldsymbol{E}_{\text{q out}}^{\parallel} - \epsilon_s \boldsymbol{E}_{\text{q in}}^{\parallel} \cdot \boldsymbol{E}_{\text{q in}}^{\parallel}\right)\hat{\boldsymbol{q}}\end{array}\right]_{r=a} \tag{5B}$$

$$-\frac{1}{2}\left[\begin{array}{l}\left(\mu_b \boldsymbol{H}_{\text{out}}^{\perp} \cdot \boldsymbol{H}_{\text{out}}^{\perp} - \mu_s \boldsymbol{H}_{\text{in}}^{\perp} \cdot \boldsymbol{H}_{\text{in}}^{\perp}\right)\hat{\boldsymbol{n}} + \left(\mu_b \boldsymbol{H}_{\text{m out}}^{\parallel} \cdot \boldsymbol{H}_{\text{m out}}^{\parallel} - \mu_s \boldsymbol{H}_{\text{m in}}^{\parallel} \cdot \boldsymbol{H}_{\text{m in}}^{\parallel}\right)\hat{\boldsymbol{m}} \\ + \left(\mu_b \boldsymbol{H}_{\text{q out}}^{\parallel} \cdot \boldsymbol{H}_{\text{q out}}^{\parallel} - \mu_s \boldsymbol{H}_{\text{q in}}^{\parallel} \cdot \boldsymbol{H}_{\text{q in}}^{\parallel}\right)\hat{\boldsymbol{q}}\end{array}\right]_{r=a}$$

Now, by adding Eq (4B) and (5B) and after doing some calculations, we get exactly the surface force law of Helmholtz, at the object boundary:

$$\boldsymbol{f}_{\text{Helmholtz}}^{\text{Surface}} = \left[-\frac{1}{2}\boldsymbol{E}^2\Delta\varepsilon - \frac{1}{2}\boldsymbol{H}^2\Delta\mu\right]_{r=a} \tag{6B}$$